\documentstyle[prl,aps]{revtex}
\begin{document}
\newcommand{\nwc}{\newcommand}
\nwc{\be}{\begin{equation}}
\nwc{\ee}{\end{equation}}
\nwc{\bea}{\begin{eqnarray}}
\nwc{\eea}{\end{eqnarray}}
\nwc{\ba}{\begin{array}}
\nwc{\ea}{\end{array}}
\nwc{\rtr}{\rangle}
\nwc{\ltr}{\langle}
\nwc{\ket}[1]{|#1\rtr}
\nwc{\bra}[1]{\ltr#1|}
\nwc{\scal}[2]{\bra{#1}#2\rtr}
\nwc{\dagg}{\mbox\footnotesize{\dag}}
\nwc{\emp}{\emphasize}
\nwc{\lb}{\label}
\nwc{\rf}[1]{~(\ref{#1})}
\nwc{\ci}[1]{~\cite{#1}}
\nwc{\pa}{\partial}
\nwc{\paf}[2]{\frac{\pa#1}{\pa#2}}
\nwc{\ra}{\rightarrow}
\nwc{\Tr}{\mbox{\rm Tr}}
\nwc{\real}{\mbox{\rm Re}}
\nwc{\im}{\mbox{\rm Im}}
\nwc{\bino}[2]{\mbox{$\left(\begin{array}{c}#1\\#2\end{array}\right)$}}
\def\a{\alpha}
\def\b{\beta}
\def\e{\epsilon}
\def\l{\lambda}
\def\m{\mu}
\def\n{\nu}
\def\f{\phi}
\def\p{\pi}
\def\t{\theta}
\def\D{\Delta}
\def\O{\Omega}
\def\om{\omega}
\def\s{\sigma}
\nwc{\eps}{\epsilon}
\nwc{\br}{\mbox{\bf{R}}}
\nwc{\bc}{\mbox{\bf{C}}}
\nwc{\cl}{{\cal L}}
\nwc{\cz}{{\cal Z}}
\nwc{\cd}{{\cal D}}
\nwc{\nh}{{\hat n}}
\nwc{\up}{\uparrow}
\nwc{\down}{\downarrow}
\nwc{\cdag}{c^{\dagg}}
\nwc{\bdag}{b^{\dagg}}
\nwc{\de}{\delta}
\nwc{\lv}{{\mbox{\boldmath $l$}}}
\nwc{\ov}{\hat{ {\mbox{\boldmath $\O$}} }}
\nwc{\fv}{\hat{ {\mbox{\boldmath $\phi$}}  }}
\nwc{\Sv}{{\mbox {\boldmath $S$}}}

\tolerance=10000

\twocolumn[\hsize\textwidth\columnwidth\hsize
     \csname @twocolumnfalse\endcsname

\title{Effective actions for spin ladders.}
\author{S. Dell'Aringa$^1$, E. Ercolessi$^{1,2}$, G. Morandi$^{1,3}$, 
P. Pieri$^{1,3}$ and M. Roncaglia$^{1,3}$\\
\begin{center}
{\small {\em $^1$ Dipartimento di Fisica, Universit\`a di Bologna,}}\\
{\small {\em Via Irnerio 46, I-40126, Bologna, Italy.}}\\
{\small {\em $^2$ INFM, Bologna, Italy.}}\\
{\small {\em $^3$ INFM and INFN, Bologna, Italy.}}\\
\end{center}}
\maketitle
\begin{abstract}
We derive a path-integral expression for the
effective action in the continuum limit of an AFM
Heisenberg spin ladder with an arbitrary number of legs.
The map is onto an $O(3)$ nonlinear $\sigma$-model
(NL$\sigma$M) with the addition of a topological term that
is effective only for odd-legged ladders and half-odd
integer spins. We derive the parameters of the effective
NL$\sigma$M and the behaviour of the spin gap for the case
of even-legged ladders.
\end{abstract}

\pacs{PACS: 74.20.Hi, 71.10.+x, 75.10.Lp}
\vskip2pc]
\newpage

Triggered by the discovery of high-T$_c$
superconductivity\ci{1} and by the fact that,
in the strong-coupling regime, the Hubbard
model maps  onto an AFM Heisenberg model\ci{2}, quantum
spin systems have become in recent years of great
theoretical and experimental interest\ci{3}.
Independently from the possible connections with
high-T$_c$ materials, some years ago Haldane\ci{hal1} put
forward the conjecture that the value of the spin should
discriminate dramatically between half-odd integer and
integer one-dimensional spin chains: the former should
be gapless, with power-law decay of spin-spin
correlations, while the latter should be characterized by
exponential decay of correlations and hence by a
nonvanishing spin gap. It turns out\ci{5} that, in the
continuum limit, an AFM Heisenberg chain is described by
an $O(3)$ nonlinear $\sigma$-model
(NL$\sigma$M)\ci{6} with the addition of a topological
term (a Pontrjagin index\ci{6}), multiplied by a
coefficient $\theta = 2\pi S$, where $S$ is the value of
the spin. As shown, e.g. by Shankar and Read\ci{sha}, it
is precisely this term that makes the half-odd integer
spin chain (where $\theta = \pi \; \mbox{mod} \; 2\pi$)
massless, while it is ineffective for integer spin chains
(where $\theta = 0  \mbox{ mod} \; 2\pi$), which are gapped.

Extensions of these results to D=2 turned out to be
rather disappointing, to the extent that it has been
proved in a convincing way [8-12] that, at least for smooth
configurations (see however\ci{hal} for the case of a
singular ``hedge-hog" field configuration), the
topological term is absent irrespective of the value of
the spin and also of the topology of the lattice.

Quite recently, however, it came as a surprise when
theoretical (mostly numerical) and subsequently
experimental results showed that spin {\it ladders},
obtained by antiferromagnetically coupling a finite
number of chains, show drastically different behaviours
that are basically determined by the value of the spin
{\em and} by the number of legs in the ladder. 
Coupled spin chains had already been studied previously \ci{sch} mainly as 
models of single chains with spin higher than the minimum value $S=1/2$. A 
systematic study of the appearance of a spin gap in even-legged ladders 
and of the possible onset of superconductivity upon doping with holes 
was initiated later on by Dagotto et al. \ci{DRS} who made for the first 
time quantitative predictions on both the spin gap and on the pair 
correlation function. Odd-legged ladders are instead gapless. 
This ``even-odd" conjecture 
was formulated for the first time by Rice et al. \ci{RGS}, and is very much 
reminiscent of the already mentioned Haldane's ``integer-half integer" 
conjecture \ci{hal1} for single spin chains. According to this ``even-odd" 
conjecture, undoped integer-spin ladders are gapped while 
half-odd integer spin
ladders are gapped when the number of legs is even,
gapless when it is odd. These results have by now strong
support, both theoretically\ci{7} and experimentally\ci{8} 
(for a recent review see\ci{9}).
The situation seems to be not so simple for conducting (i.e. hole-doped) 
ladders. A different behaviour between even and odd-legged ladders is 
predicted only for very strong intrachain repulsions \ci{Schu}.
On the contrary, in the weak coupling regime, analytical \ci{Schu,AR} and 
numerical \ci{KKA} studies suggest that d-wave interchain pairing correlations 
become dominant over AF fluctuations in both the two and the three-legged 
ladder, due to the simultaneous presence of gapless and gapped spin modes. 
So, the situation is richer than in the case of single conducting chains, 
that display a Luttinger liquid type \ci{9,S2} behaviour. 

Following mainly the ideas of Haldane\ci{hal1} and
Affleck\ci{5} on the role of the topological term in
spin chains, considerable theoretical effort has been
recently devoted to the investigation of the existence
and the role of such a term when a finite number of
chains are coupled to form a ladder (the possible topological origin
for the different behaviour of even and odd leg ladders has been
suggested for the first time in\ci{dz}). In particular
S\'en\'echal\ci{sen} has given a derivation of the 
NL$\sigma$M continuum limit of a two-leg ladder using a
coherent-state path-integral\ci{11} expression for the
partition function, while Sierra\ci{sie} has employed
an (operator) Hamiltonian approach following closely
Affleck's mapping of the Heisenberg chain onto the 
NL$\sigma$M.

What we present here is a path-integral
approach that is however different from S\'en\'echal's.
It allows us to reproduce S\'en\'echal's results for the
two-leg ladder, but also to generalize them to ladders
with an arbitrary number of legs and to obtain a clear
understanding of why the topological term is absent for
even-legged ladders.

The Hamiltonian for a ladder system with $n_l$ legs of lenght $N$ is
defined by:
\bea
& &H=\sum_{a=1}^{n_l}\sum_{i=1}^{N}
\left\{ J_a \Sv_a(i)\cdot\Sv_a(i+1)\right.\nonumber\\
&+& \left. J_{a,a+1}'\Sv_a(i)\cdot\Sv_{a+1}(i)
\right\}\; .
\lb{hl}
\eea
The only condition we shall impose on the coupling costant $J_a$ 
and $J'_{a,a+1}$ is that the classical minimum of the 
Hamiltonianian\rf{hl} be antiferromagnetically ordered. 
The partition function for the Hamiltonian\rf{hl} in a path-integral 
representation which makes use of spin coherent states\ci{11} is given by:
\be
\cz(\b)=\int [D \ov]\exp\left\{ i s \sum_{i,a}\om[\ov_a(i,\tau)]
-\int_0^{\beta}d\tau H(\tau)\right\}\; ,
\ee  
where $\om[\ov_a(i,\tau)]$ is the Berry phase factor coming from the 
exponentiation of the overlap between coherent states at nearby 
timeslices\ci{aue}, while $H(\tau)$ is obtained by replacing
in the Hamiltonian\rf{hl} the operator $\Sv_a(i)$ by the classical variable 
$s\ov_a(i,\tau)$. Explicitly: 
\bea
\om[\ov(\tau)]&=&\int_0^\b d\tau\, {\dot \phi}(\tau)(1-\cos\t(\tau))\\
H(\tau)&=&\sum_{a,i}
\left\{ J_a s^2\ov_a(i,\tau)\cdot\ov_a(i+1,\tau)\right.\nonumber\\
&+& \left. J_{a,a+1}'s^2\ov_a(i,\tau)
\cdot\ov_{a+1}(i,\tau)\right\}\; .
\eea
The Berry phase measures the area enclosed by the path 
$\ov(\tau)=(\sin \t(\tau)\cos\f(\tau),\sin\t(\tau)\sin\f(\tau),
\cos\t(\tau))$ on the unit sphere.   
We shall proceed now along the lines of Ref.\ci{hal}. We assume that 
short-range antiferromagnetic correlations survive at the quantum level, so that 
we consider the dominant contribution to the path-integral as coming from 
paths described by:
\be
\ov_a(i,\tau)=(-1)^{a+i}\fv(i,\tau) \left(1-\frac{|\lv_a(i,\tau)|^2}{s^2}
\right)^{1/2} \!\! + 
\frac{\lv_a(i,\tau)}{s}\; .
\lb{an}
\ee
The fluctuation field $\lv_a(i)$ is supposed to be small: $|\lv_a(i)/s|<<1$
and the field $\fv(i,\tau)$ slowly varying. We are then allowed to make 
an expansion up to quadratic order in $\lv$, $\fv'$ and $\dot {\fv}$.
The parametrization of the classical fields in\rf{an} is different from
that employed in S{\'e}n{\'e}chal's approach\ci{sen} to the two-leg ladder, and 
allows for generalization of the path-integral approach to ladders 
with an arbitrary number of legs. In particular, the fact that the field
$\fv$ is chosen to depend on the site index $i$ along the legs, but not 
on the index $a$ labeling the sites along the rung, reflects the 
assumption that the correlation length be much greater 
than the total width of the ladder, i.e. $\xi >> n_l a$, where $\xi$ is the 
staggered spin-spin correlation length and $a$ the lattice spacing.
This is well supported
by several numerical works (see for instance\ci{cor} and\ci{whi}). 
The constraint $\ov_a^2(i)=1$ implies $\fv^2(i)=1$ and $\fv(i)\cdot\lv_a(i)=0$.
As far as the intra-leg term of $H(\tau)$ is concerned, everything 
proceeds as for the continuum limit of the 1-dimensional Heisenberg chain
(see\ci{aue} for a detailed calculation) and hence we write just the 
final result:
\bea
& &\sum_{a,i}
J_a s^2\ov_a(i,\tau)\cdot\ov_a(i+1,\tau)\approx\nonumber\\
& & \int dx \left(\frac{(s^2\sum_aJ_a)}{2}\fv^{'2}(x,\tau)
+ 2 \sum_a J_a |\lv_a(x,\tau)|^2\right)\; .
\eea
For the inter-leg term we have instead:
\bea
& &\ov_a(i,\tau)\cdot\ov_{a+1}(i,\tau)\approx\nonumber\\
& &-1 +\frac{|\lv_a(i,\tau)|^2}{2s^2}+\frac{|\lv_{a+1}(i,\tau)|^2}{2s^2}
+\frac{\lv_a(i,\tau)\cdot\lv_{a+1}(i,\tau)}{s^2}\, .
\eea
The term $H(\tau)$ in the action has then the continuum limit:
\bea
& &H(\tau)=\nonumber\\
& &\frac{1}{2}\int dx\left[ (s^2\sum_aJ_a)\fv^{'2}(x,\tau)
+ \sum_{a,b}\lv_a(x,\tau)L_{a,b}\lv_b(x,\tau)\right]\; ,
\nonumber
\eea
where $L_{a,b}$ is the same matrix defined in Ref.\ci{sie}, 
i.e.:
\be
L_{a,b}=\cases{4J_a+J'_{a,a+1}+J'_{a,a-1}&$a=b$\cr
               J'_{a,b}&$|a-b|=1$}
\ee
($J'_{a,a-1}\equiv J'_{a-1,a}$ and $J'_{1,0}=J'_{n_l,n_{l+1}}=0$ 
in the formula above). We have finally to 
evaluate the Berry phase term. We now need the formula for the variation 
of the Berry phase $\om[\fv]$ upon a small change $\delta \fv$\ci{aue}:
\be
\delta \om=\int_0^{\b}d\tau\,\delta \fv \cdot (\fv\times \dot {\fv})\; .
\ee
We have therefore, at leading order:
\bea
& &\sum_{i,a}\om[\ov_a(i,\tau)]=s\sum_{i,a}(-1)^{a+i}\om[\fv(i,\tau)]
\nonumber\\
&+&\sum_{a,i}\int_0^{\beta}d\tau\, (\fv(i,\tau)\times \dot{\fv}(i,\tau))
\cdot \lv_a(i,\tau)\nonumber\\
&\equiv&\Gamma[\fv]
+\sum_{a,i}\int_0^{\beta}d\tau\, (\fv(i,\tau)\times \dot{\fv}(i,\tau))
\cdot \lv_a(i,\tau)\; .
\lb{top}
\eea 
The first term in the last equation is the topological term. We shall 
return to it in a while, after having integrated out the fluctuation 
field $\lv_a(i,\tau)$:
\bea
& &
\int D[\lv]\exp\int_0^{\b}d\tau\int dx\left[
-\frac{1}{2}\sum_{a,b}\lv_a(x,\tau)L_{a,b}\lv_b(x,\tau)\right.\nonumber\\
&+&\left. i
\sum_a\left(\fv(x,\tau)\times\dot{\fv}(x,\tau)\right)\cdot \lv_a(x,\tau)\right]
\nonumber\\
& &=\exp \left[-\int_0^{\b}d\tau\int dx\,\frac{\sum_{a,b}L_{a,b}^{-1}}{2}
|\fv\times\dot{\fv}|^2\right] \nonumber\\
& &=\exp \left[-\int_0^{\b}d\tau\int dx\,\frac{\sum_{a,b}L_{a,b}^{-1}}{2}
|\dot{\fv}|^2\right]\; .
\eea
The integration over the field $\lv_a(i)$ gives therefore the 
kinetic term of the NL$\s$M. For the topological term:
\be
\Gamma[\fv]=s\sum_{a}(-1)^{a}\sum_i(-1)^i\om[\fv(i,\tau)]\; ,
\ee
we can simply observe that, since the field $\fv(i,\tau)$ does not
depend on $a$, one has just the same topological term as for the chain
times a factor which is zero for even $n_l$ and 1 for odd $n_l$:
\be
\Gamma[\fv]=\cases{\frac{\theta}{4\pi}\int_0^{\b}d\tau\int dx\,
\fv\cdot(\dot{\fv}\times\fv')&$n_l$ odd\cr
0&$n_l$ even}
\ee
($\theta=2\pi s$ in the last equation).
In the case of the 2D (infinite) lattice, cancellation of the 
topological term results, as is well known [8-12], from taking the 
continuum limit also along the direction of the rungs, and holds at 
least for smooth field configurations. Here it is instead a direct 
consequence of the assumption that $\xi>>n_l a$, which lies at the heart of 
the parametrization chosen in Eq.\rf{an}. The staggering of the field 
${\hat \phi}$ along the rungs leads then to the result that the topological 
term survives only for odd-legged ladders, being of course significant 
only for half-odd integer spins.
Putting everything together we found that the antiferromagnetic Heisenberg 
ladder system is mapped onto a (1+1) NL$\s$M with the euclidean Lagrangian:
\be
\cl=\cases{\frac{1}{2g}\left(\frac{1}{v_s}\dot{\fv}^2+v_s{\fv}^{'2}
\right) 
+\frac{i\theta}{4\pi}\fv\cdot(\dot{\fv}\times\fv')&$n_l$ odd\cr
\frac{1}{2g}\left(\frac{1}{v_s}\dot{\fv}^2+v_s{\fv}^{'2}\right)
&$n_l$ even}\; ,
\ee
where the NL$\s$M parameters are defined by:
\bea
g^{-1}&=&s\left(\sum_{a,b,c}J_aL^{-1}_{b,c}\right)^{1/2}\\
v_s&=&s\left(\frac{\sum_a J_a}{\sum_{b,c}L^{-1}_{b,c}}\right)^{1/2}
\; .
\eea
Let us remark that the NL$\s$M velocity we have obtained coincides with the 
spin-wave velocity (see\ci{sie} for the calculation of the spin-wave 
velocity in our model Hamiltonian). In the particular case $n_l=2$ we 
obtain:
\bea
g&=&\frac{1}{s}(1+J'/2J)^{1/2}\\
v_s&=&2sJ(1+J'/2J)^{1/2}\; .
\eea
These values of the parameters concide with the ones obtained 
by S{\'e}n{\'e}chal in\ci{sen} for the two-leg ladder.
They are different, however, from the parameters obtained by Sierra through 
his Hamiltonian mapping to the NL$\s$M\ci{sie}.
He has for the $n_l$-leg ladder:
\bea
g^{-1}&=& s\left[2\sum_{a,b,c}J_aL^{-1}_{b,c}
-\frac{1}{4}\delta_{n_l}\right]^{1/2}\\
v_s&=&s\left[2\frac{\sum_aJ_a}{\sum_{b,c}L^{-1}_{b,c}}-\delta_{n_l}
\frac{1}{\left(2\sum_{b,c}L_{b,c}^{-1}\right)^2}\right]^{1/2}\; ,
\eea
where $\delta_{n_l}$ is equal to 1 for odd $n_l$ and 0 for even $n_l$.
For even $n_l$ his $g$ is smaller, while $v_s$ is greater, by a factor of 
$\sqrt{2}$ as compared with ours. The two results coincide (and coincide
with Afflecks's as well, as they should) in the limit of a single chain, 
i.e for $n_l=1$. They are again different for odd $n_l>1$. We argue that 
this is due to the introduction, in Ref.\ci{sie}, of additional massive 
fields. The latter are decoupled from the $\sigma$-model field and are 
ultimately neglected in the effective theory developed in\ci{sie}. 
Nonetheless, they appear to affect the actual values of the parameters 
of the effective $\sigma$-model. The agreement of our approach and of 
that of Ref.\ci{sie} for $n_l=1$ (when there is no room for massive fields) 
seems to give support to our conjecture.

In any case the general picture for the behaviour
of the spin-ladder is the same as in\ci{sie}. The ladders with an
even number of legs are mapped to a (1+1) NL$\s$M without  
topological term and are therefore gapped, while the ones with an odd
number of legs are described by a (1+1) NL$\s$M with the topological 
term, and are gapless\ci{sha} for half-odd integer spin.
In the case $J'_{a,a+1}=J'$, $J_a=J$ and
$J'<<J$ we have $g\sim 2/sn_l$; the coupling costant $g$ gets therefore
smaller and smaller when the number of legs increases and the NL$\s$M 
enters in a weak coupling regime.  In this regime we may use the 
formula $\Delta\sim \exp(-2\pi/g)$\ci{pol} to estimate the spin gap 
for even-leg ladders. 
We obtain then $\Delta\sim \exp(-\pi s n_l)$; the gap decreases therefore with 
the number of legs, as observed in numerical simulations\ci{cor,whi} and as 
expected from the fact that in the limit $n_l\to\infty$ the difference between
ladders with even and odd number of legs must disappear.
Let us also remark that the ratio 
$\xi/n_la\sim\exp(\pi s n_l)/n_l$; the condition 
$\xi<<n_l a$ we supposed at the beginning of our calculation seems 
therefore to be satisfied (in a self-consistent way) better and better when the
number of legs increases (at least for $J'<<J$).
In the opposite regime $J'>>J$ the coupling costant
becomes strong since $g\sim (J'/J)^{1/2}$ and we can therefore estimate 
the spin gap  as $\Delta=v_sg$ in the strong coupling regime\ci{sha}. 
For two legs we get $\Delta\sim J'$ when $J'>>J$, a result 
which also agrees with what found in literature through other techniques\ci{9}.
\vskip1cm
{\bf Remark.} Just after completion of this work a  preprint by G. Sierra has 
appeared\ci{pre}, in which the mapping of spin ladders to the NL$\s$M is 
reviewed also within a path-integral formalism. The values for the NL$\s$M
parameters found by Sierra agree with ours.
\vskip1cm
We wish to thank M.A. Martin-Delgado and G. Sierra for interesting us in the 
problem. One of us (P.P.) is grateful to E. Arrigoni, F.D.M. Haldane and
D.V. Khveshchenko for useful discussions during the Euroconference
``Correlations in Unconventional Quantum Liquids", Evora 5-11 October
1996, where he was supported by an EC Grant.

\end{document}